\documentclass[useAMS]{mn2e}

\usepackage{amssymb,amsmath,psfig,times}
\title[Flaring activity of IGR J22517+2217]
{Modeling the flaring activity of the high z, hard X-ray selected blazar IGR J22517+2217 }
\author[Lanzuisi et al.]
{G. Lanzuisi$^1$, A. De Rosa$^1$, G. Ghisellini$^2$, P. Ubertini$^1$, 
\newauthor{F. Panessa$^1$, 
M Ajello$^3$, L. Bassani$^4$, Y. Fukazawa$^5$, F. D'Ammando$^6$}
\\
\\
$^1$IASF--Roma/INAF, Via del Fosso del Cavaliere 100, I-00133 Roma\\
$^2$INAF -- Osservatorio Astronomico di Brera, via E. Bianchi 46, I–23807 Merate, Italy\\
$^3$SLAC National Laboratory and Kavli Institute
for Particle Astrophysics and Cosmology, 2575 Sand Hill Road, Menlo Park,
CA 94025, USA\\
$^4$IASF--Bologna/INAF, Via Gobetti 101, I-4012, Bologna, Italy\\
$^5$Department of Physical Science, Hiroshima University, 1-3-1 Kagamiyama, Higashi-Hiroshima, Hiroshima 739-8526, Japan\\
$^6$INAF -- Istituto di Astrofisica Spaziale e Fisica Cosmica, Via U. La
Malfa 153, I-90146 Palermo, Italy\\
}

\begin{document}


\maketitle

\begin{abstract}

We present new {\it Suzaku} and {\it Fermi} data,
 and re-analyzed archival hard X-ray data from INTEGRAL and {\it Swift}--BAT survey, 
to investigate the physical properties of 
the luminous, high-redshift, hard X-ray selected blazar 
IGR J22517+2217, through the modelization of its broad band spectral energy distribution (SED)
in two different activity states.
Through the analysis of the new {\it Suzaku} data and the flux selected data from archival hard X-ray  observations,
we build the source SED in two different states, one for the newly discovered flare occurred in 
2005 and one for the following quiescent period.
Both SEDs are strongly dominated by the high energy hump peaked at 10$^{20}$-10$^{22}$ Hz, that is at least two  
orders of magnitude higher than the low energy (synchrotron) one at 10$^{11}$-10$^{14}$ Hz, and 
varies by a factor of 10 between the two states. 
In both states the high energy hump is modeled as inverse Compton emission 
between relativistic electrons and seed photons produced externally to the jet, while
the synchrotron self--Compton component is found to be negligible.
In our model the observed variability can be accounted for by a variation of the total number
of emitting electrons, and by a dissipation region radius changing from within to outside
the broad line region as the luminosity increases.
In its flaring activity, IGR J22517+2217 shows one of the most powerful jet
among the population of extreme, hard X--ray selected, high redshift blazar observed so far.

\end{abstract}


\section{Introduction}

In the past years, {\it Integral}--IBIS (Ubertini et al. 2003)
and {\it Swift}--BAT (Barthelmy et al. 2005) catalogs of hard X--ray selected 
extragalactic sources have been published (Sambruna et al. 2007; Bird et al. 2010), opening 
a new window in the study of blazars.

The hard-X-ray selection produces a set of diverse sources, including extremely ``red'' 
flat spectrum radio quasars (FSRQs), with Inverse Compton (IC) peak at keV-MeV
energies, or ``blue'' BL Lacs with synchrotron peak at these frequencies,
and provides a way to test the validity of the blazar sequence, 
looking for outliers not predicted by the sequence itself (Giommi et al. 2007).
In particular, by selecting hard X--ray luminous FSRQ at high redshift, i.e. with the highest intrinsic 
bolometric luminosities, it is possible to collect samples of 
supermassive black holes (SMBHs) in the early Universe, thus introducing important observational 
constraint on the number density of 
heavy black holes at high redshift and hence on their formation processes and timing.

IGR J22517+2217 was first reported by Krivonos et al.
(2007) as an unidentified object detected by  {\it Integral}--IBIS (150 ks of total IBIS exposure).
{\it Swift} follow-up observations were used to associate the source 
with MG3 J225155+2217 (Bassani et al. 2007). MG3 J225155+2217 has been optically identified as a QSO by Falco et al. (1998) in a redshift survey of 
177 FSRQs, on the basis of  S$_{IV}$, Ly$\alpha$, C$_{II}$ and C$_{IV}$ emission lines.
IGR J22517+2217 is the highest redshift ($z=3.668$) blazar detected in the fourth {\it Integral}--IBIS hard X--rays catalog (Bird et al. 2010).
The source has a {\it Swift}--BAT counterpart (SWIFT J2252.0+2218) in the 3 years BAT survey (Ajello et al. 2009) and is present in the multifrequency ``Roma--BZCAT" 
catalog of blazars (Massaro et al. 2009).

Using XRT, IBIS and archival optical/IR data, Bassani et al. (2007) constructed a non simultaneous SED 
of the source, showing an extremely bright X--ray emission with respect to the
optical emission ($\alpha_{OX}<0.75$), 
and suggested that IGR J22517+2217 could be a rare FSRQ with synchrotron peak at X--ray frequencies,
or a more canonical FSRQ, i.e. with the synchrotron peak at radio--mm frequencies and IC peak at MeV-GeV energies, 
but with an exceptionally strong Compton dominance.

This ``controversial" blazar has been studied also by Maraschi et al. (2008). 
They reanalyzed the existing {\it Swift} (XRT and BAT) and {\it Integral}--IBIS data, 
and propose a ''standard leptonic one-zone emission model'' (Ghisellini \& Tavecchio 2009, see Sect. 3) with the peak of the synchrotron component at microwave/radio frequencies, 
and a high luminosity external Compton (EC) component peaking in hard X-rays to reproduce the SED  of the source.
This model ruled out both a synchrotron and a synchrotron self--Compton (SSC) interpretation for the 
X--ray emission. 

Ghisellini et al. (2010) included IGR J22517+2217 in their sample of 10 X--ray selected blazar at $z>2$:
the intent of the paper was to characterize the physical properties of these powerful sources, and to confirm
the capability of the hard X--ray selection in finding 
extreme blazars with massive SMBH, powerful jets and luminous accretion disks. 
IGR J22517+2217 is the highest redshift FSRQ in their sample and shows the highest total jet power (P$_{Jet}=1.5\times10^{48}$ erg s$^{-1}$).

All these previous studies have been performed through the analysis of the ''average'' X-ray spectra obtained with the INTEGRAL and {\it Swift}-BAT surveys, 
without taking into account any possible flux variation of the source during the period of monitoring (5 years for INTEGRAL and BAT).
In this paper we present the discovery of strong flaring activity, 
in X--ray (IBIS and BAT) archival data, of this extremely bright and peculiar
FSRQ, and the modelization of both its flaring and quiescent SEDs.
New {\it Suzaku} and {\it Fermi} data are used for characterizing the  
quiescent state.
Our goal is to investigate the evolution 
of the SED and obtain information on the physical condition of the source in the two different states.

The paper is organized as follows. In \S2 we report the multiwavelength data analysis of the instruments involved in the SED building. 
In \S3 we describe the model adopted to reproduce the broad band SED, while in \S4 we discuss the SED fitting of both the flaring and quiescent state. 
The summary of our results is presented in \S5.
Throughout the paper, a $\Lambda-CDM$ cosmology with $H_0 = 71$ km s$^{-1}$ Mpc$^{-1}$, 
$\Omega_\Lambda = 0.73$, and $\Omega_m = 0.27$ is adopted.

\section{Data reduction}

\subsection{ New {\it Suzaku} data}

{\it Suzaku} observed the source on 2009 Nov 26 (ID 704060010, PI A. De Rosa), 
for a net exposure of $\sim$40 ks, with both the X--ray Imaging Spectrometer (XIS; Koyama et al. 2007) and the 
Hard X--ray Detector (HXD; Takahashi et al. 2007).
The XIS instrument has 3 operating detectors at the time of the observation: the front illuminated XIS 0 and XIS 3, sensitive in the 0.5--10 keV band, and the 
back--illuminated XIS 1, extending the low energy range to 0.2 keV.
The HXD is instead composed of GSO scintillators and silicon PIN diodes.
The PIN detectors observe in the 12--60 keV energy band, while the GSO ones can observe up to 600 keV.
Data reduction and processing were performed using HEASOFT $v6.9$ and {\it Suzaku ftools v16}.
Only XIS and PIN data have been used in this analysis, since the source is below the 
sensitivity limit of the GSO scintillators.
The cleaned event files produced by the data processing with the standard selection criteria were used.

XIS source events were extracted from a region of radius 200 arcsec, and the 
background events from an annulus (external radius 400 arcsec) outside the source region.
The response matrix and effective area were calculated for each detector using {\it Suzaku} 
ftools tasks {\it xisrmfgen} and {\it xissimarfgen}.
Given that XIS 0 and XIS 3 have similar responses, their event files were summed.
Only data in the 0.5--8 keV band were considered, and the spectral counts were 
rebinned using at least 30 counts per bin, in order to allow the use of $\chi^2$ statistic.

PIN data were extracted from the HXD cleaned event files after standard screening. 
The tuned background model supplied
by the {\it Suzaku} team (Fukazawa et al. 2009), was used for the ``Non X--ray Background" (NXB) events. 
The background light curve was corrected for the 10x oversampling rate. 
The source spectra were corrected for deadtime using {\it hxddtcor}. 
We estimate the cosmic X--ray background (CXB) contribution to the
PIN background using the model given in Gruber et al. (1999),
which is folded with the PIN response to estimate the CXB rate.

The XIS and PIN data were fitted with a simple power-law, modified by neutral absorption at the source redshift, plus 
galactic absorption, fixed to the value measured by Kalberla et al. (2005) at the source 
coordinates\footnote{http://heasarc.nasa.gov/cgi-bin/Tools/w3nh/w3nh.pl} ($N_{H, Gal.} = 5\times10^{20} cm^{-2}$). 
The best fit values are reported in Table 1.
The XIS and PIN data show a flat spectrum, with $\Gamma=1.5\pm0.1$ and $\Gamma=1.5\pm0.8$ 
respectively, with the PIN power law normalization being $\sim1.2$ times the XIS one, a known cross calibration issue
\footnote{http://heasarc.gsfs.nasa.gov/docs/suzaku/analysis/abc sec 5.4}.
The XIS data show some curvature below 1 keV, that can be reproduced either by an intrinsic 
column density of $N_H=(1.5\pm1.1)\times10^{22}$ cm$^{-2}$ at the source redshift (compatible with the value found in Bassani et al. using XRT data)
or with a broken power-law with break energy of $0.8\pm0.1$ keV and $\Delta\Gamma=0.8$.
Both models give a comparable reduced $\chi^2$ of 1.03 and 1.04, respectively, thus the 
quality of the data do not allow do disentangle the two possibilities.
In Sect. 4 we will attempt to disentangle these two different models using broadband data. 

The hard X-ray flux obtained by fitting the PIN data is a factor of $\sim$10 (4) lower that the one
reported in literature from {\it Integral}-IBIS ({\it Swift}-BAT) survey data (Bassani et al. 2007; Maraschi et al. 2008; Ghisellini et al. 2010).
This suggests that the source was in a much less active state during the 2009 {\it Suzaku} observation, compared to previous measurements.
This lead us to reanalyze the archival IBIS and BAT data in order to check for the presence of variability in the hard X-ray source flux.
In Figure 1 top panel we show the {\it Suzaku} XIS (red, black) and PIN (green) unfolded spectra and model best fit found for IGR J22517+2217.  
For comparison also the IBIS spectrum (blue) is shown, to emphasize the different flux state.

\subsection{{\it Integral}-IBIS and {\it Swift}-BAT survey data}

\begin{figure}
\begin{center}
\psfig{figure=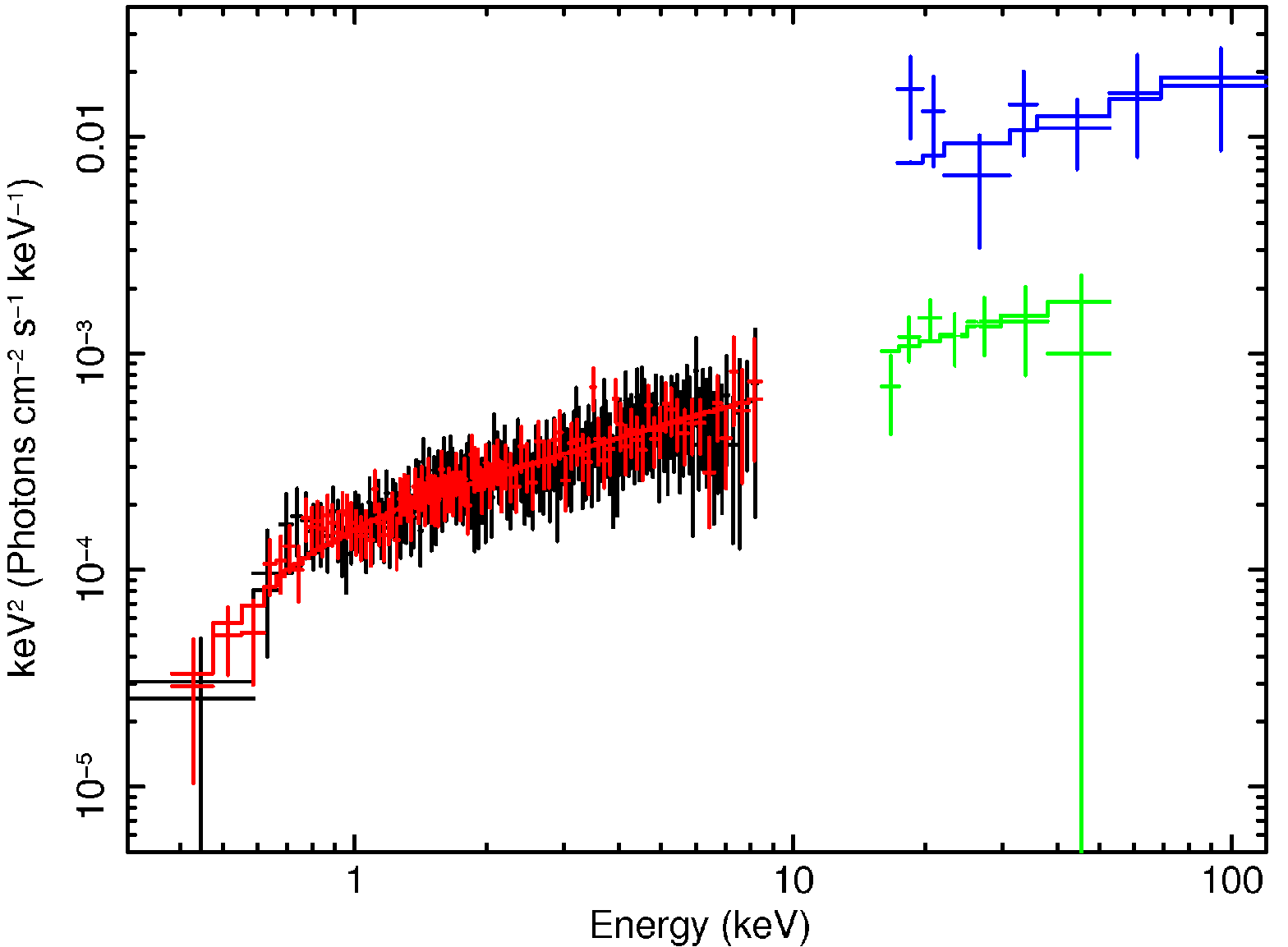,width=8cm,height=6.5cm}\hspace{1cm}\psfig{figure=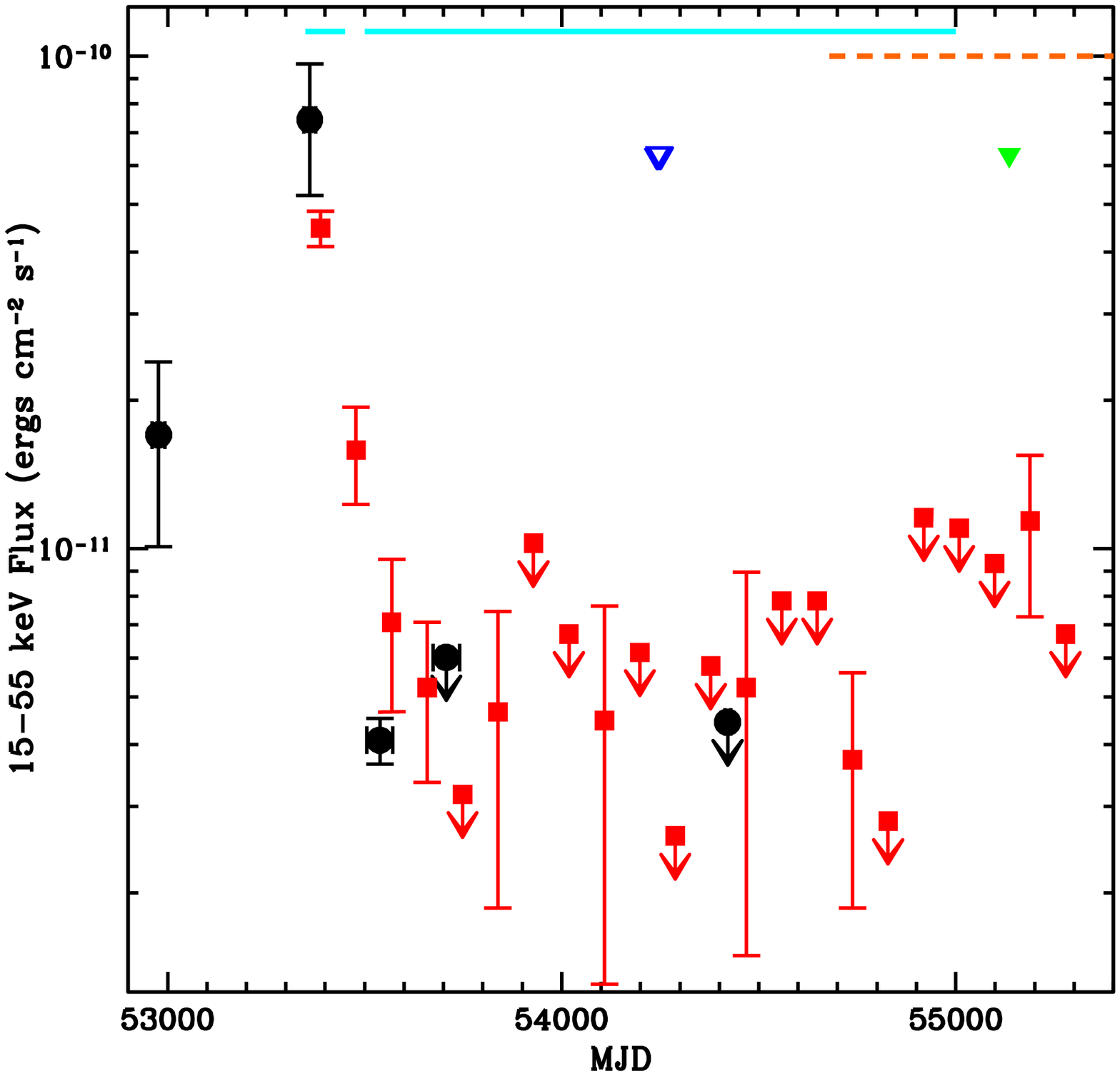,width=8cm,height=7cm}
\caption{
{\it a) Top panel:} {\it Suzaku} XIS (red, black) and PIN (green) unfolded spectrum and model of IGR J22517+2217. 
For comparison the IBIS cat4 spectrum is shown (blue). 
{\it b) Bottom panel:} Hard X--ray light curve of IGR J22517+2217. 
Red squares (black circles) represent BAT (IBIS) 15--55 keV flux. 
The blue empty (green filled) triangles show the time of XRT/UVOT ({\it Suzaku}) observation. 
The cyan solid lines represent the time intervals used in the BAT spectra extraction.
The orange dashed line represents the period of observation of {\it Fermi}/LAT.
}
\end{center}
\label{fig:light}
\end{figure}

IBIS data are taken from the Fourth IBIS/ISGRI Soft $\gamma$--ray Survey Catalog (Bird et al. 2010), 
covering the period Feb 2003 -- Apr 2008.
Images from the ISGRI detector (Lebrun et al. 2003) for each pointing have been generated
using scientific Analysis
Software (Goldwurm et al. 2003) OSA version 7.0. 
Five primary energy bands (20--40, 30--60, 20--100, 17--30, and
18--60 keV) were used to maximize the detection sensitivity
for sources with various energy spectra (for details see Bird et al. 2010).

The BAT analysis results presented in this paper were derived with all the available data 
during the time interval 2005 Jan 19 -- 2010 Apr 14.
The 15--55 keV spectra and light curve were extracted following the recipes presented in Ajello 
et al. (2008, 2009). 
The spectra are constructed by weighted averaging of the source spectra extracted from short 
exposures (e.g., 300s) and are accurate to the mCrab level. 
The reader is referred to Ajello et al. (2009) for more details. 

The total IBIS and BAT spectra can be reproduced with a simple power law having $\Gamma=1.6\pm 0.6$ and 
$1.6\pm0.5$, respectively;
the 15--55 keV flux is F$_{15-55 \, {\rm keV}}=(2.5\pm0.9)\times10^{-11}$ and  
F$_{15-55 \, {\rm keV}}=(2.1\pm0.8)\times10^{-11}$ erg cm$^{-2}$ s$^{-1}$, respectively.

The bottom panel of Fig. 1 shows the BAT (red squares) and IBIS (black diamonds) historical light curve 
of IGR J22517+2217 from 2003 Dec 04 to 2010 Mar 23.
The 15--55 keV IBIS light curve was extracted from the ISDC Science Products Archive\footnote{http://www.isdc.unige.ch/heavens\_webapp/integral} 
adopting a nominal bin size of 100 days. 
We converted the observed IBIS and BAT count rates into 15--55 keV observed flux, using the 
WebPimms HEASARC tool\footnote{http://heasarc.nasa.gov/Tools/w3pimms.html}, 
assuming as underling model a power law with photon index $\Gamma=1.6$, consistent with 
the values observed in the BAT and IBIS spectra, and assuming a constant cross-calibration between IBIS and BAT, equal to one.

\noindent Fig.1 (bottom panel) shows that: 

-- the source displays quite strong long term variability in hard X--rays;

-- a strong flare episode occurred around Jan 2005, and the source reached a 15--55 keV flux maximum of $(8\pm2) \times10^{-11}$ erg cm$^{-2}$ s$^{-1}$
(a factor of 20 higher than the flux measured by {\it Suzaku}-PIN in 2009); 

-- after the flare, the source faded into a quiescent state, reaching a flux that is at or below the detection limit of both BAT and IBIS instruments.
As can be seen, the IBIS light curve is completely dominated by the flare, 
i.e. the source flux is below the IBIS detection limit after 
MJD 53550 and the total spectrum extracted from the entire period can be considered representative of the flare.

As a result of its different pointing strategy, BAT has a much more regular and extended coverage of the source, 
and we were able to characterize the source in both states.
We extracted a BAT spectrum from the period around the flare of 2005 (2004 Dec 11--2005 Mar 21)
and also one from the remaining quiescent period (2005 May 10--2009 Jun 18, 
solid cyan lines in the bottom panel of Fig. 1).

The BAT flux relative to the flare state is the highest one measured in the hard X-ray energy range
(F$_{15-55 \, {\rm keV}}=(3.7\pm0.8)\times10^{-11}$ erg cm$^{-2}$ s$^{-1}$).
During this state the source is detected up to 200 keV, and the spectrum is characterized by photon index of 1.5$\pm$0.5.
In the quiescent state the source is detected, with significance $\sim3\sigma$, only up to $\sim75$ keV, and 
the spectrum has a flux F$_{15-55 \, {\rm keV}}$ a factor of $\sim15$ lower than the flaring one, while the photon index is 1.7$\pm$1.1.
Considering the large uncertainties on $\Gamma$ we can consider the spectra (in flaring and quiescent state) comparable (see
Table 1 for IBIS and BAT spectral analysis results).

\begin{table*}
\begin{center}
\label{tab:xray}
\begin{tabular}{ccccccccccccccc}
\hline\hline\\
\multicolumn{1}{c} {Date}&
\multicolumn{1}{c} {Inst.}&
\multicolumn{1}{c} {Exp.}&
\multicolumn{1}{c} {$N_{\rm H}$}&
\multicolumn{1}{c} {$\Gamma$}&
\multicolumn{1}{c} {F$_{2-10}$}&
\multicolumn{1}{c} {$\log L_{2-10}$}&
\multicolumn{1}{c} {F$_{15-55}$}& 
\multicolumn{1}{c} {$\log L_{15-55}$}\\
 (1) & (2) &(3) & (4) & (5) & (6) & (7) & (8) & (9) \\
\hline\\  
2003 Dec 04--2007 Nov 17    & IBIS       & 191 & -                & 1.6$\pm$0.6  & -          & -    &25.1$\pm$8.9 & 48.1 \\
2004 Dec 11--2005 Mar 21    & BAT Flare  & 160 & -                & 1.5$\pm$0.5  & -          & -    &36.9$\pm$8.1 & 48.3 \\
2005 May 10--2009 Jun 18    & BAT quiesc.&1610 & -                & 1.7$\pm$1.1  & -          & -    &2.6$\pm$1.6  & 47.3 \\
2007 May 01                 & XRT        & 40 & $2.0\pm1.5$       & 1.4$\pm$0.1  & 2.4$\pm$0.4& 47.1 & -           & -     \\
2009 Nov 01                 & XIS        & 40 & $1.5\pm1.1$       & 1.5$\pm$0.1  & 1.2$\pm$0.1& 46.7 & -           & -    \\
2009 Nov 01                 & PIN        & 40 & -                 & 1.5$\pm$0.8  & -          & -    &3.8$\pm$1.8  &  47.4\\
\hline      
\hline                              
\end{tabular}
\end{center} 
\caption{Best fit model parameters  for the different observations of IGR 22517+2217 analyzed in this work.
The broadband continuum is reproduced with a simple power-law, modified by intrinsic absorption at the source redshift, 
plus galactic absorption N$_{\rm H, Gal}$ = 5$\times$10$^{20}$ cm$^{-2}$.
Column: 
(1) Date;
(2) Instrument; 
(3) Effective exposure (ks);
(4) Column density (in $10^{22}$ cm$^{-2}$ units);
(5) Photon index; 
(6) 2--10 keV observed flux (in $10^{-12}$ erg s$^{-1}$ cm$^{-2}$ units);       
(7) Log 2--10 keV deabsorbed luminosity (erg s$^{-1}$);
(8) 15--55 keV flux (in $10^{-12}$ erg s$^{-1}$ cm$^{-2}$ units);
(9) Log 15--55 keV luminosity (erg s$^{-1}$).}
\end{table*}

\subsection{{\it Fermi}/LAT data}

{\it Fermi} Large Area Telescope (LAT; Atwood et al. 2009) data were collected from Aug 2008 (MJD 54679) to Aug 2010 (MJD 55409). 
During this period, the {\it Fermi}/LAT instrument operated mostly in survey mode, 
scanning the entire $\gamma$-ray sky every 3 hours.
The analysis was performed with the ScienceTools software package version {\it v9r17p0}, which
is available from the Fermi Science Support Center. 
Only events having a high probability of being photons
-- those in the ``diffuse class" -- were used. 

The energy range used was 100 MeV -- 100 GeV, and the maximum zenith angle value was $105^{\circ}$.
We adopted the unbinned likelihood analysis, using the standard P6\_V3\_DIFFUSE response functions.
The diffuse isotropic background used is {\it isotropic\_iem\_v02} and the galactic diffuse emission model used is 
{\it gll\_iem\_v02}\footnote{http://fermi.gsfc.nasa.gov/ssc/data/access/lat/BackgroundModels.html}.
We considered a region of interest (RoI) of $15^{\circ}$ from the IGR J22517+2217 position.
All sources from the 1FGL catalog (Abdo et al. 2010) within the RoI of the source were included in the fit, 
with their photon indexes and the integral fluxes free to vary, 
plus a point source with power law spectrum at the IGR J22517+2217 position, having photon index fixed to 2.2 (a value typical of FSRQ in the GeV band).
The normalization of the background was also left free to vary.
We also repeated the analysis using the source list of the 2FGL catalog (Ackermann et al. 2011), obtaining consistent results.

IGR J22517+2217 is located $\sim6^{\circ}$ from 3C 454.3, a bright, extremely variable source in 
the $\gamma$--ray sky (Ackermann et al. 2010). It contributes for more than 90\% to the total counts in the RoI.
We tried to exclude the period of flaring activity of 3c 454.3 from the data set, in order to minimize the contamination by this source.
The results obtained for IGR J22517+2217, however, did not change significantly.

IGR J22517+2217 is not detected in the 2 year observation, and the computed Test Statistic (TS, Mattox et al. 1996)
is $TS \simeq3.5$ in the full band.
We therefore calculated the 95\% upper limits in 5 energy bands (i.e. 0.1--0.3, 0.3--1, 1--3, 3--10, 10--100 GeV), using the profile likelihood method.
The upper-limits were corrected for attenuation due to extragalactic background light (EBL) through $\gamma-\gamma$ interactions (Chen et al. 2004, Razzaque et al. 2009), 
although only the 10-100 GeV band was found to be affected by significant attenuation.
As a consequence of being close to 3C 454.3, the background around the position 
of IGR J22517+2217 is higher than in a typical extragalactic field, making it difficult to constrain more strongly the upper limits  at GeV energies.
The upper limits are plotted in Figure 2, with all other data discussed in this paper.

Given that {\it Fermi} data were collected starting from Aug 2008, they completely fall
in the quiescent period of the source, and so it is not surprising that IGR J22517+2217 was not detected.
Therefore, they will be used to characterize the 
state of the source in the quiescent SED only, while we do not have available $\gamma$-ray data 
for the flaring state SED.

\subsection{Archival Data}

We collected archival radio and optical data from NASA/IPAC EXTRAGALACTIC DATABASE \footnote{http://ned.ipac.caltech.edu/} for the source.
Radio data at 1.4 4.8 and 8.4 GHz comes from NRAO/VLA.
Optical data in J, H, and K bands are taken with the  UFTI instrument at UKIRT (Kuhn 2004).
The UV data are taken from Ghisellini et al. (2010), where they corrected the {\it Swift}-UVOT observed magnitudes
for the absorption of neutral hydrogen in intervening Lyman$-\alpha$ absorption systems. 
We also reanalyzed the archival XRT spectrum (average of 4 contiguous observations), obtaining consistent results with those found in Bassani et al. (2007)
using the same set of data, and with the results reported in \S2.1 for the XIS spectra.
The only difference is in the XRT observed 2-10 keV flux, that is a factor 2 higher than the flux measured by XIS.

\begin{figure*}
\begin{center}
\psfig{figure=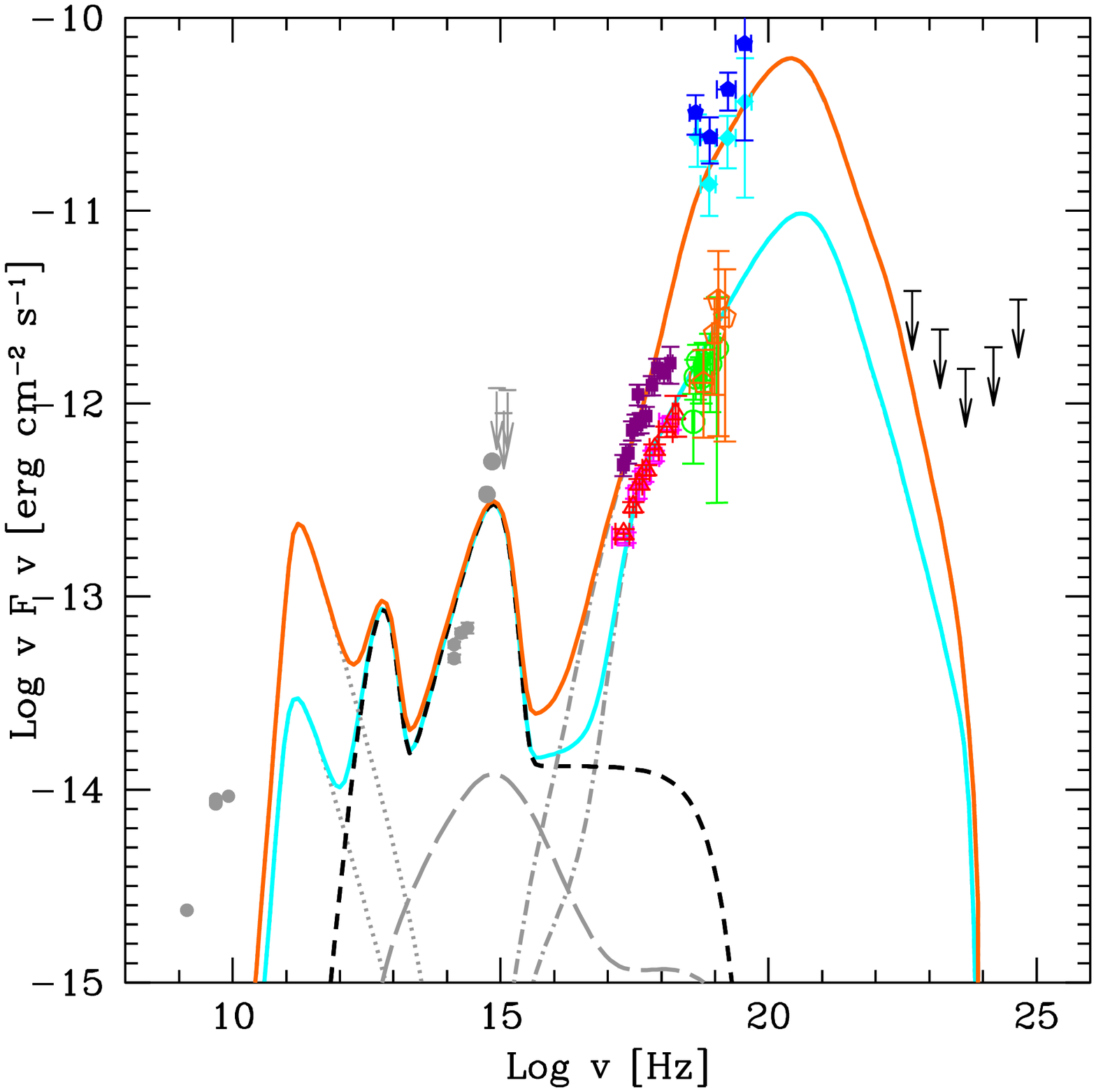,width=16cm,height=12cm}
\hspace{0.3cm}
\vspace{-0.5cm}
\psfig{figure=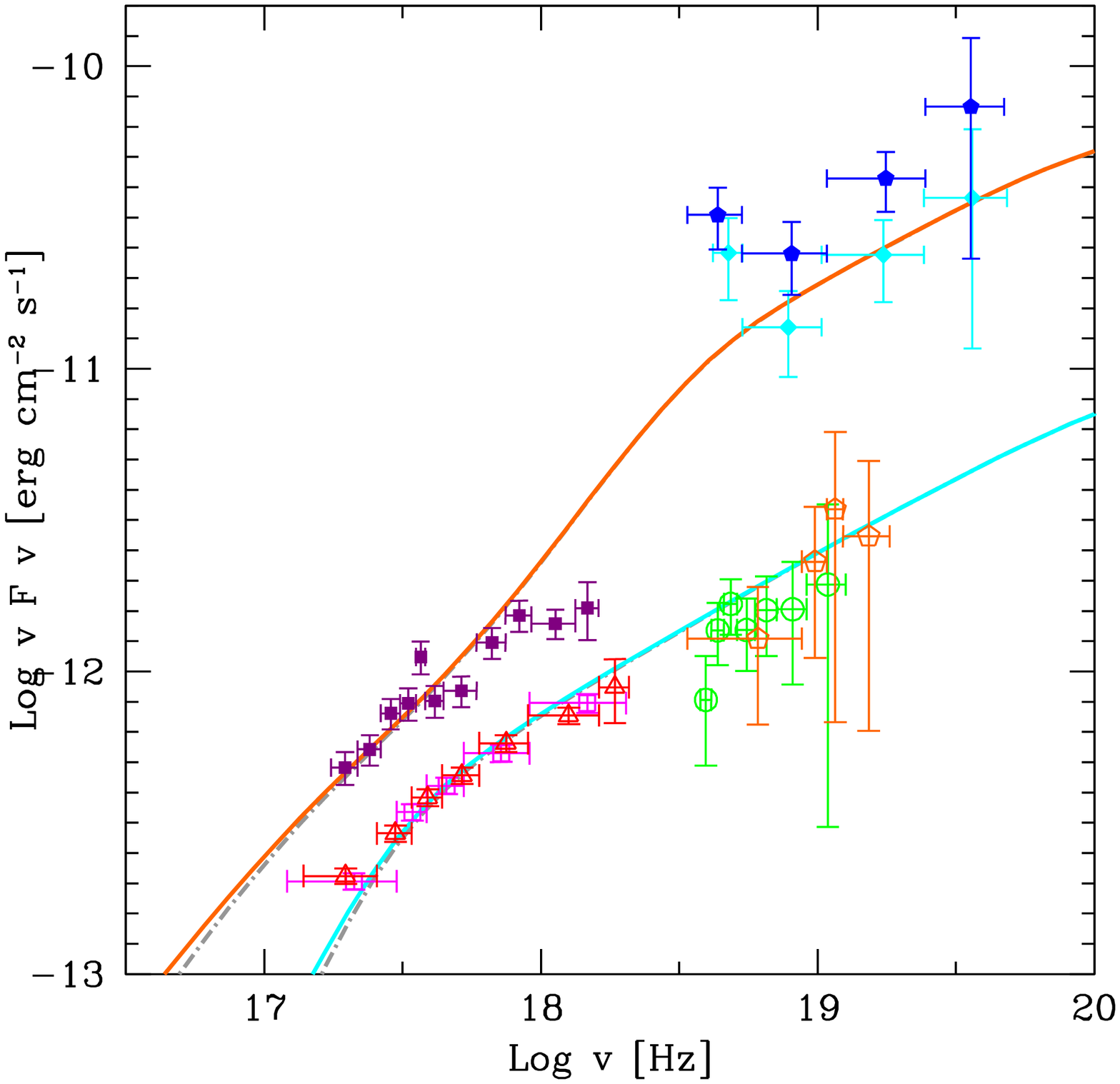,width=10cm,height=8cm}
\caption{
{\it a) Top panel:} 
Spectral energy distribution of IGR J22517+2217. 
Gray circles and arrows represent archival radio/optical/UV data from NED. 
Empty red triangles and magenta  
squares represent XIS 0 and XIS 3 data, 
empty green circles and orange pentagons represent PIN and BAT quiescent data respectively, 
while black arrows are {\it Fermi} upper limits in 5 bands.
Filled violet squares represent XRT data, filled cyan diamonds and blue pentagons represent 
IBIS and BAT flare data respectively.
The solid cyan and orange curves are the results of the modeling of the  quiescent 
and flaring states, respectively.
With gray lines we show the different components of the non--thermal emission:
synchrotron (dotted), synchrotron self--Compton (long dashed) and
external Compton (dot--dashed).
The black dashed line corresponds to the thermal emission of the disk, the IR torus and the X--ray disk corona. 
The model does not account for radio emission, produced from much larger regions of the jet.
{\it a) Bottom panel:} Zoom in the X--ray energy range for the two SEDs. Symbols as in top panel.
}
\end{center}
\label{fig:sed2}
\end{figure*}

\section{The SED model}

The model adopted to fit the SED is a leptonic, one--zone synchrotron and inverse Compton model, 
fully discussed in Ghisellini \& Tavecchio (2009).
The assumptions can be summarized as follows:

\begin{table*}
\label{tab:sed}
\begin{center}
\begin{tabular}{ccccccccc}
\hline\hline\\
\multicolumn{1}{c} {State}&
\multicolumn{1}{c} {$R_{\rm diss}$}&
\multicolumn{1}{c} {$P'_{\rm inj}$}&
\multicolumn{1}{c} {$B$}&
\multicolumn{1}{c} {$\Gamma$}&
\multicolumn{1}{c} {$\gamma_{\rm b}$}&
\multicolumn{1}{c} {$\gamma_{\rm max}$}&
\multicolumn{1}{c} {$s_1$}&
\multicolumn{1}{c} {$s_2$}\\
 (1) & (2) &(3) & (4) & (5) & (6) & (7) & (8) & (9)  \\
\hline\\ 
Low  & 570 (1900)  &   0.045   & 1.06  & 16 & 70 & 2e3 & -1  & 4  \\
High & 990 (3300)  &   0.30    & 0.61  & 16* & 70* & 2e3* & -1*  & 4*  \\
\hline                                    
\end{tabular}\end{center}     
\caption{Input parameters of the SED fitting for the low and high state of IGR J22517+2217.
Column: 
(1) State ; 
(2) dissipation radius in units of $10^{15}$ cm and, in parenthesis, in units of
Schwarzschild radii;
(3) intrinsic injected power ($10^{45}$ erg s$^{-1}$) in the form of relativistic electrons;
(4) magnetic field intensity (Gauss);
(5) bulk Lorentz factor at R$_{diss}$; 
(6) and (7) break and maximum random Lorentz factors of the injected electrons; 
(8) and (9) slopes of the injected electron distribution [Q($\gamma$)] below and above $\gamma_{\rm b}$;
* fixed values from low to high state SED.}
\end{table*}

-- The emitting region is assumed to be spherical 
(with radius $R$) and at a distance $R_{\rm diss}$ (dissipation radius) from the central black hole.
The emitting electrons are injected at a rate Q($\gamma$) [cm$^{-3}$ s$^{-1}$] for a finite time equal to the
light crossing time $R/c$. 
The adopted function Q($\gamma$) is a smoothly broken power law with a break at $\gamma_{\rm b}$
and slopes $s_1$ and $s_2$ below and above $\gamma_{\rm b}$, respectively.
The emitting region is moving with a velocity $\beta_{\rm c}$ corresponding to a bulk Lorentz factor $\Gamma$. 
We observe the source at the viewing angle $\theta_{\rm v}$.

-- The external radiation sources taken into account are:
the broad line region (BLR) photons, assumed to re--emit 10\% of the 
accretion disk luminosity from a shell--like distribution of clouds located at a distance 
$R_{\rm BLR}=10^{17}L_{\rm d, 45}^{1/2}$ cm (Kaspi et al. 2005), where $L_{\rm d}$ is the disk luminosity; 
the IR emission from a dusty torus located at a distance 
$R_{\rm IR}=2.5\times10^{18}L_{d, 45}^{1/2}$ cm (Elitzur 2006); 
the direct emission from the accretion disk, including its X--ray corona. 
Also the starlight contribution from the inner region of the host galaxy and the cosmic
background radiation are taken into account, but these photon sources are unimportant in
the case of IGR J22517+2217.

-- The accretion disk is a standard ``Shakura \& Syunyaev" (1973) disk, emitting as a blackbody at each radius.
The maximum temperature ($T_{\rm max}$), i.e. the peak of the disk luminosity, is assumed to occur at $\sim5$ 
Schwarzschild radii ($R_S$).
Thus from the position of the peak of the disk luminosity, and the total luminosity of the accretion 
disk ($L_d$) it is possible to derive $M_{\rm BH}$ and $\dot M$, once a value for the efficiency $\eta$ is assumed ($\eta=0.08$ for a Schwarzschild black hole). 
See Ghisellini et al. (2010a) for a discussion on caveats of this black hole mass estimate.

We can estimate the black hole mass $M_{\rm BH}$ and the accretion luminosity 
$L_{\rm d}$ of IGR J22517+2217 using optical and UV data and upper-limits corrected for the absorption
of neutral hydrogen in intervening Lyman $\alpha$ systems along the line of sight 
(see Ghisellini et al. 2010a for details).
Given the uncertainties in the amount of intervening Ly$\alpha$ systems 
and the paucity of data the results must be considered an approximation.
We find $M_{\rm BH}=10^9 M_{\sun}$ and $L_{\rm d}=6.8\times10^{46}$ erg s$^{-1}$.
These values correspond to a disk radiating at 45\% of the Eddington level.

The BLR are located at $R_{\rm BLR}= 8\times 10^{17}$ cm and the 
IR emitting torus at $R_{\rm IR}=2.5\times10^{19}$ cm.
The total X--ray corona luminosity is assumed to emit 30\% of $L_{\rm d}$. 
Its spectral shape is assumed to be $\propto \nu^{-1} \exp(-h\nu/150 \, {\rm keV})$.

\section{Discussion}

Our analysis shows that the extremely Compton dominated FSRQ IGR J22517+2217,
experienced a strong flare in the high energy hump in Jan 2005, and then faded in a quiescent state.
In order to investigate the physical properties of the source we built two SEDs for the two different states, and fitted 
the data with the leptonic, one--zone synchrotron and inverse Compton model described in \S3.

In order to build the SED for the quiescent state of IGR J22517+2217, we used the  X-ray data from {\it Suzaku} (XIS and PIN), 
the BAT spectrum extracted from the quiescent period and {\it Fermi}/LAT 24 months upper limits.
The archival non--simultaneous optical/UV data were added to our data.

For the flaring SED we used hard X-ray data from {\it Integral}--IBIS and the {\it Swift}--BAT 
spectrum extracted from the Jan 2005 flare.
We do not have soft X-ray data available for the flaring period. 
However, in order to put some constraint at soft X-ray frequencies we chose to include the XRT spectrum in 
the flaring SED. This has a factor 2 higher normalization with respect to the XIS data.
We stress that, as observed in other samples of bright, red blazars (Ghisellini et al. 2010a), 
the flux variability is larger at higher energies (i.e. in hard X--rays)
but modest at few keV. This support our choice of include the XRT data in the flaring SED.

All data points and SED model components of the flaring and quiescent SED are plotted in Fig. 2, top panel. Fig. 2, bottom
panel, shows a zoom in the region of X-ray data.
The strong and very bright hard X-ray spectrum, together with the upper limits in the {\it Fermi}/LAT
energy range, constrain the peak of the high energy hump of the quiescent SED to be located 
at $\sim10^{20}$--$10^{21}$ Hz, and thus the form of energy spectrum of radiating electrons.
The same energy spectrum is then assumed in the flaring SED, for which no MeV/GeV data are available.
The corresponding synchrotron peak falls in a region of the spectrum localized around 
$10^{11}$ Hz for both SEDs.

However, in our single--zone leptonic model, the synchrotron and the high 
energy humps are produced by the same population of electrons.
Furthermore, if the high energy emission is given by the external Compton process,
the energies of the electrons emitting at the $\sim$MeV peak are rather modest,
implying a corresponding low frequency synchrotron peak.
We then require that the synchrotron component peaks at low energies,
close to the self--absorption frequency, and furthermore require that the
thin synchrotron emission has a steep spectrum, whose flux is smaller
than the optical archival data (characterized instead by a rather hard slope,
that we interpret as emission from the accretion disk).

Given the relative paucity of the observational data, the choice of the model parameters is not unique,
and some further assumption has to be made.
We indeed fix the viewing angle $\theta_{\rm v}$ to 3$^\circ$, to be close to the
$\theta_{\rm v}\sim 1/\Gamma$ condition.

Note that we model both states of
the source by changing the minimum number of parameters, given that fitting 
two data sets with the same model, allows to constraint better
the model parameters.
In particular, we assume that the accretion luminosity does not change 
for the two states, and require also the bulk Lorentz factor and the parameters
of the distribution of the injected electrons (break and maximum random Lorentz factors
and slopes below and above $\gamma_{\rm b}$) to be the same for the two SEDs.
Thus, the parameters that are left free to vary from one state to the other are 
the dissipation radius $R_{\rm diss}$, the injected power $P'_{\rm inj}$ and the 
magnetic field $B$, that is proportional to $1/R_{\rm diss}$ (the Poynting flux is assumed to be constant).

The results of our modeling are shown in Fig. 2, where we show the total flux together
with the contributions of the non--thermal (synchrotron, self Compton, external Compton)
and thermal (accretion disk, IR torus, X--ray corona) components.
As can be seen, the curvature around 1 keV observed in the XRT and XIS spectra
can be well reproduced by an EC component changing (softening) slope from 
$\sim10^{17}$ to $\sim10^{19}$ Hz, disfavoring the intrinsic obscuration scenario
for the shape of the X-ray emission of IGR J22517+2217.

Table 2 reports the parameters of the SED fitting for the quiescent and flaring states. 
The main difference between the two SEDs is the power $P'_{\rm inj}$ 
injected in the source in the form of relativistic electrons, that changes by a factor $\sim$7.
The increase of $P'_{\rm inj}$ accounts for the enhanced X--ray flux in the high state.
The other difference is the location of the emitting region $R_{\rm diss}$, 
becoming larger for the high state.
This is dictated by the detailed modeling of the slope of the soft to hard X--ray spectrum,
requiring an electron distribution with a break at low energies, and another break
at somewhat larger energies.
This is accounted for, in our modeling, by requiring that electron of very low energies
(corresponding to random Lorentz factors $\gamma<5$) do not cool in one light crossing time.
This can be achieved if the location of the emitting region is slightly beyond the BLR,
in a zone where the BLR radiation energy density is somewhat smaller.
This is the reason leading to a larger $R_{\rm diss}$ in the high state.
As a consequence of assuming a larger region, the magnetic field is lower,
following the assumption of a constant Poynting flux ($\propto B^2 R^2$).
The large Compton dominance constrains the value of the magnetic field,
and in turn the relevance of the self Compton flux, found to be almost negligible.

This is also in agreement with the results pointed out in Sikora et al. (2009):
bright blazars with very hard ($\alpha_x<0.5$) X--ray spectra and high luminosity ratio 
between high and low frequency spectral components
challenge both the standard synchrotron self-Compton and 
hadronic models, while EC can easily account for these observed properties.

In the analysis described above, the bulk Lorentz factor is assumed to be constant. 
However, the change of bulk Lorentz factor is often invoked to explain the variability of FSRQs.
As a further check, we performed a new fit of both low and high states, 
leaving $\Gamma$ as a free parameter, in addition to $R_{\rm diss}$ and $P'_{\rm inj}$, although the fit is limited by the poor number of data points

Both SEDs are well reproduced with these new parameters for the low (high) state: $\Gamma$=15 (20); $R_{\rm diss}$=1700 (3500) $R_S$ and 
Log($P'_{\rm inj}$)= 43.48 (44.17) erg s$^{-1}$.
In this case the values of $P'_{\rm inj}$ are slightly lower for both low and high states,
and the difference is lower (a factor of $\sim5$), while the difference 
in $R_{\rm diss}$ is slightly larger: 
from 1700 to 3500 $R_S$ instead of 1900 and 3300 $R_S$.
Thus the change in $\Gamma$ has the main effect of slightly decreasing the required variation of the total injected power
to account for the observed variability, but no other substantial differences are introduced.
This new fit also gives an idea of the degree of degeneracy in the fit parameters, due to the incompleteness of the data set,
especially in the $\gamma$-ray band.

In Table 3 we report the logarithm of the jet power in the form of radiation, 
Poynting flux and bulk motion of electrons and protons, 
in erg s$^{-1}$, calculated for both SEDs.
They have been calculated from
\begin{equation}
P_{\rm i} = \pi R^2 \Gamma^2 c U^\prime_{\rm i} 
\end{equation}
where $U^\prime_{\rm i}$ is the energy density of interest, calculated in the comoving frame
(see e.g. Celotti \& Ghisellini 2008).
As discussed in Ghisellini et al (2011), the power $P_{\rm r}$ dissipated by the jet
to produce the radiation we see is almost model--independent, since it depends
only on the observed luminosity and on the bulk Lorentz factor $\Gamma$.
The power dissipated in other forms, on the other hand, depends on the amount of electrons ($P_{\rm e}$),
protons ($P_{\rm p}$), and magnetic field ($P_{\rm B}$) carried by the jet,
which have been estimated by applying our specific model.
Furthermore the power carried in the bulk motion of protons requires 
a knowledge of how many protons are there, per emitting lepton.
The values given in Table 3 assume one proton for one emitting lepton.

\begin{table}
\label{tab:sed}
\begin{center}
\begin{tabular}{cccccccccccccccc}
\hline\hline\\
\multicolumn{1}{c} {State}&
\multicolumn{1}{c} {$\log P_{\rm r}$}&
\multicolumn{1}{c} {$\log P_{\rm B}$}&
\multicolumn{1}{c} {$\log P_{\rm e}$}&
\multicolumn{1}{c} {$\log P_{\rm p}$}&\\
 (1) & (2) &(3) & (4) & (5) \\
\hline\\ 
Low  & 46.04 & 45.54 & 45.00 & 47.56 \\
High & 46.83 & 45.54 & 46.17 & 48.41 \\
\hline                                    
\end{tabular}\end{center}   
\caption{Input parameters of the SED model fitting both the low and high state of IGR J22517+2217.      
Column: 
(1) State ; 
(2)--(5) logarithm of the jet power in the form of radiation ($P_{\rm r}$), 
Poynting flux ($P_{\rm B}$), bulk motion of electrons ($P_{\rm e}$) and protons ($P_{\rm p}$,
assuming one proton per emitting electron). Powers are in erg s$^{-1}$.}
\end{table}

From Table 3 we can see that the power
$P_{\rm r}$ changes from $\sim0.15\times L_{\rm d}$ in the quiescent state, 
to $P_{\rm r} \sim L_{\rm d}$ in the flaring state, i.e. the jet requires a power comparable 
to the disk luminosity to produce the radiation we see, in the latter case.

Tanaka et al. (2011) reported a similar behavior, based on {\it Fermi}-LAT data, for the strong 2010 GeV flare of the blazar 4C+21.35:
assuming similar efficiencies for the accretion disk and the jet, they estimated a jet intrinsic power $L_{jet}$
changing from $\sim0.1 L_{acc}$ in the quiescent state to $\sim1 L_{acc}$ (beeing $L_{acc}$ the intrinsic accretion power).
They argued that these results, combined with the findings of Abdo et al. (2010b) on several FSRQ detected by {\it Fermi}-LAT,
suggest a scenario in which the observed $\gamma$-ray variability of blazars is due to the different power of the jet,
that normally represents only a small fraction of the accretion power, while during major flares is capable of 
carry away almost all the available accretion power.
Our findings are quantitatively similar, and thus result in agreement with this view.

The $P_{\rm r}$ however is a {\it lower limit} to the total jet power.
The total jet power, dominated by the bulk motion of protons associated to emitting 
electrons, is $P_{\rm jet} = P_{\rm B} + P_{\rm e} + P_{\rm p} = 3.6\times10^{47}$ 
and $2.6\times10^{48}$ erg s$^{-1}$, in the low and high state, respectively. 
$P_{\rm jet}$ is dominated by $P_{\rm p}$: if there is indeed one proton per
emitting lepton, then the jet power is from 3 to 30 times more powerful than the 
accretion luminosity. 

It has been proposed that jets in luminous blazars may well be numerically dominated by pairs,
but being still dynamically dominated by protons (Sikora \& Madejski 2000, Kataoka et al. 2008).
We computed the jet power due to protons assuming an upper limit for this ratio of one proton per 20 pairs.
Above this limit the jet is too ``light'' and the Compton drag produces significant
deceleration (Ghisellini \& Tavecchio 2010).
The one-to-one ratio can be assumed as a lower limit for this ratio.

The lower limits for $P_{\rm p}$ obtained in this way are $1.8\times10^{46}$ and $1.3\times10^{47}$ erg s$^{-1}$ for the low and high state, respectively.
Therefore with this assumption, the jet power in electron, protons and magnetic field, becomes comparable with the radiation power.
This translates into a total jet power of $P_{\rm jet} = 2.2\times10^{46}$  and $1.4\times10^{47}$ erg s$^{-1}$, respectively.

The values obtained assuming one proton per lepton are extreme, even if compared with the distribution of $P_{\rm jet}$ and $L_{\rm d}$ 
computed, with the same assumptions, for a sample of high redshift {\it Fermi}/LAT and BAT blazars in Ghisellini et al. (2011, Fig. 9), 
that are in the range $P_{\rm jet}\simeq10^{46}-2\times10^{48}$ and $L_{\rm d}\simeq8\times10^{45}-2\times10^{47}$ erg s$^{-1}$. 
To better clarify the remarkable behavior of IGR J22517+2217, 
we note that similar values of jet power have been achieved during the exceptional flare 
of 3C 454.3 in December 2009 (Ackermann et al. 2010; Bonnoli et al. 2011). Therefore, IGR J22517+2217 represents one of the ''monsters'' of the high-z Universe.

\section{Summary and conclusion}

Thanks to a new {\it Suzaku} observation in the X-ray energy band, the {\it Fermi} upper limits in the 0.1-100 GeV band,
the flux selected spectra obtained through a re-analysis  of the IBIS and BAT hard X-ray data, and other optical and radio archival data sets, 
we were able to identify a strong flare episode in the high redshift, hard X-ray selected blazar IGR J22517+2217, which occurred in Jan 2005,
followed by a period of quiescence, extending up to the present days.
To model the overall SEDs of the source in the flare and quiescent states, we adopted a leptonic, one-zone synchrotron and inverse Compton model. 
The optical/UV emission is interpreted as thermal emission from the accretion disk, plus IC from the corona, and reprocessed emission from the BLR. 

The curvature observed in the X-ray spectra was proposed to be due to intrinsic, moderate absorption (N$_H\sim2\times10^{22}$ cm$^{-2}$).
However, in the context of the broad band SED modelization proposed in this paper, it appears to be inherently accounted for by an intrinsic softening of the 
EC component around $\sim10^{18}$ Hz.   

In both states a very strong Compton dominance is observed, with the high energy hump (produced by EC), that is at least two orders of magnitudes 
higher than the low energy (synchrotron) one.
The high energy peak flux varies by a factor 10 between the two states, while the high energy peak frequency remain almost constant, between $10^{20}-10^{22}$ Hz.
The observed large Compton dominance constrains the value of the magnetic field,
and hence the relevance of the self--Compton component, that is found to be negligible in both states.
The model can explain the observed variability as a variation of the total number of emitting electrons (a variation of factor $\sim7$) 
and as a change in the dissipation radius, moving from within to outside the broad line region as the luminosity increases.

In the flaring state, the jet power lower limit, represented by the radiative component $P_{\rm r}$,
requires a power comparable to the disk luminosity to produce the observed radiation.
The total jet power upper limit, dominated by the bulk motion of protons, and estimated assuming one proton per electron, 
is more than $\sim30$ times more powerful than the accretion luminosity ($2.6\times10^{48}$ erg s$^{-1}$). 
Such extreme values have been derived only recently for a handful of extreme, high redshift, hard-X/soft-$\gamma$ ray selected FSRQs, 
showing similar strong Compton dominance,
and comparable with the value achieved by 3C 454.3 during the 2009 exceptional flare.

\section*{acknowledgements}

We thank the referee for useful comments that improved the paper.

Partial support from the Italian Space Agency (contracts ASI-INAF
ASI/INAF/I/009/10/0) is acknowledged. 

This research has made use of the NASA/IPAC Extragalactic Database (NED) 
which is operated by the Jet Propulsion Laboratory, California Institute of Technology, 
under contract with the National Aeronautics and Space Administration.

The \textit{Fermi} LAT Collaboration acknowledges generous ongoing support
from a number of agencies and institutes that have supported both the
development and the operation of the LAT as well as scientific data analysis.
These include the National Aeronautics and Space Administration and the
Department of Energy in the United States, the Commissariat \`a l'Energie Atomique
and the Centre National de la Recherche Scientifique / Institut National de Physique
Nucl\'eaire et de Physique des Particules in France, the Agenzia Spaziale Italiana
and the Istituto Nazionale di Fisica Nucleare in Italy, the Ministry of Education,
Culture, Sports, Science and Technology (MEXT), High Energy Accelerator Research
Organization (KEK) and Japan Aerospace Exploration Agency (JAXA) in Japan, and
the K.~A.~Wallenberg Foundation, the Swedish Research Council and the
Swedish National Space Board in Sweden.

Additional support for science analysis during the operations phase is gratefully
acknowledged from the Istituto Nazionale di Astrofisica in Italy and the Centre National d'\'Etudes Spatiales in France.

\begin{appendix}


\end{appendix}


\begin{thebibliography}{}

\bibitem[Abdo et al.(2010)]{2010ApJS..188..405A} Abdo, A.~A., et al.\ 2010, ApJS, 188, 405 
\bibitem[Abdo et al.(2010)]{2010ApJ...716..835A} Abdo, A.~A., Ackermann, 
M., Ajello, M., et al.\ 2010, ApJ, 716, 835 
\bibitem[Ackermann et al.(2010)]{2010ApJ...721.1383A} Ackermann, M., et al.\ 2010, ApJ, 721, 1383 
\bibitem[The Fermi-LAT Collaboration(2011)]{2011arXiv1108.1435} Ackermann, M., et al.\ 2011, arXiv:1108.1435 
\bibitem[Ajello et al.(2008)]{2008ApJ...678..102A} Ajello, M., Greiner, J., 
Kanbach, G., Rau, A., Strong, A.~W., \& Kennea, J.~A.\ 2008, ApJ, 678, 102 
\bibitem[Ajello et al.(2009)]{2009ApJ...699..603A} Ajello, M., et al.\ 
2009, ApJ, 699, 603 
\bibitem[Barthelmy et al.(2005)]{2005SSRv..120..143B} Barthelmy, S.~D., et 
al.\ 2005, Space Sci. Rev., 120, 143 
\bibitem[Bassani et al.(2007)]{2007ApJ...669L...1B} Bassani, L., et al.\ 
2007, ApJl, 669, L1 
\bibitem[Bird et al.(2010)]{2010ApJS..186....1B} Bird, A.~J., et al.\ 2010, 
ApJS, 186, 1 
\bibitem[Bonnoli et al.(2011)]{2011MNRAS.410..368B} Bonnoli, G., 
Ghisellini, G., Foschini, L., Tavecchio, F., 
\& Ghirlanda, G.\ 2011, MNRAS, 410, 368 
\bibitem[Burrows et al.(2005)]{2005SSRv..120..165B} Burrows, D.~N., et al.\ 2005, Space Sci. Rev., 120, 165 
\bibitem[Celotti \& Ghisellini(2008)]{2008MNRAS.385..283C} Celotti, A., \& Ghisellini, G.\ 2008, MNRAS, 385, 283 
\bibitem[Chen et al.(2004)]{2004ApJ...608..686C} Chen, A., Reyes, L.~C., \& Ritz, S.\ 2004, ApJ, 608, 686 
\bibitem[Elitzur \& Shlosman(2006)]{2006ApJ...648L.101E} Elitzur, M., \& Shlosman, I.\ 2006, ApJl, 648, L101 
\bibitem[Falco et al.(1998)]{1998ApJ...494...47F} Falco, E.~E., Kochanek, 
C.~S., \& Munoz, J.~A.\ 1998, ApJ, 494, 47 
\bibitem[Fukazawa et al.(2009)]{2009PASJ...61S..17F} Fukazawa, Y., et al.\ 2009, PASJ, 61, 17 
\bibitem[Ghisellini \& Tavecchio(2009)]{2009MNRAS.397..985G} Ghisellini, G., \& Tavecchio, F.\ 2009, MNRAS, 397, 985 
\bibitem[Ghisellini et al.(2010)]{2010MNRAS.405..387G} Ghisellini, G., et al.\ 2010, MNRAS, 405, 387 
\bibitem[Ghisellini \& Tavecchio(2010)]{2010MNRAS.409L..79G} Ghisellini, G., \& Tavecchio, F.\ 2010, MNRAS, 409, L79 
\bibitem[Ghisellini et al.(2011)]{2011MNRAS.411..901G} Ghisellini, G., et al.\ 2011, MNRAS, 411, 901 
\bibitem[Giommi et al.(2007)]{2007A&A...468...97G} Giommi, P., et al.\ 2007, A\&A, 468, 97 
\bibitem[Goldwurm et al.(2003)]{2003A&A...411L.223G} Goldwurm, A., et al.\ 2003, A\&A, 411, L223 
\bibitem[Gruber et al.(1999)]{1999ApJ...520..124G} Gruber, D.~E., Matteson, 
J.~L., Peterson, L.~E., \& Jung, G.~V.\ 1999, ApJ, 520, 124 
\bibitem[Kalberla et al.(2005)]{2005A&A...440..775K} Kalberla, P.~M.~W., Burton, W.~B., Hartmann, D., Arnal, E.~M., Bajaja, E., Morras, R., Poeppel, W.~G.~L.\ 2005, A\&A, 440, 775 
\bibitem[Kaspi et al.(2005)]{2005ApJ...629...61K} Kaspi, S., Maoz, D., Netzer, H., Peterson, B.~M., Vestergaard, M., \& Jannuzi, B.~T.\ 2005, ApJ, 629, 61 
\bibitem[Kataoka et al.(2008)]{2008ApJ...672..787K} Kataoka, J., Madejski, G., Sikora, M., et al.\ 2008, ApJ, 672, 787 
\bibitem[Koyama et al.(2007)]{2007PASJ...59S..23K} Koyama, K., et al.\ 2007, PASJ, 59, 23 
\bibitem[Krivonos et al.(2007)]{2007A&A...475..775K} Krivonos, R., Revnivtsev, M., Lutovinov, A., Sazonov, S., Churazov, E., \& Sunyaev, R.\ 2007, A\&A, 475, 775 
\bibitem[Kuhn(2004)]{2004MNRAS.348..647K} Kuhn, O.~P.\ 2004, MNRAS, 348, 647 
\bibitem[Lebrun et al.(2003)]{2003A&A...411L.141L} Lebrun, F., et al.\ 2003, A\&A, 411, L141 
\bibitem[Maraschi et al.(2008)]{2008MNRAS.391.1981M} Maraschi, L., 
Foschini, L., Ghisellini, G., Tavecchio, F., 
\& Sambruna, R.~M.\ 2008, MNRAS, 391, 1981 
\bibitem[Massaro et al.(2009)]{2009A&A...495..691M} Massaro, E., Giommi, P., Leto, C., Marchegiani, P., Maselli, A., Perri, M., Piranomonte, S., \& Sclavi, S.\ 2009, A\&A, 495, 691 
\bibitem[Mattox et al.(1996)]{1996ApJ...461..396M} Mattox, J.~R., et al.\ 1996, ApJ, 461, 396 
\bibitem[Razzaque et al.(2009)]{2009ApJ...697..483R} Razzaque, S., Dermer, C.~D., \& Finke, J.~D.\ 2009, ApJ, 697, 483 
\bibitem[Sambruna et al.(2007)]{2007ApJ...669..884S} Sambruna, R.~M., 
Tavecchio, F., Ghisellini, G., Donato, D., Holland, S.~T., Markwardt, 
C.~B., Tueller, J., \& Mushotzky, R.~F.\ 2007, ApJ, 669, 884 
\bibitem[Shakura \& Sunyaev(1973)]{1973A&A....24..337S} Shakura, N.~I., \& Sunyaev, R.~A.\ 1973, A\&A, 24, 337 
\bibitem[Sikora \& Madejski(2000)]{2000ApJ...534..109S} Sikora, M., \& Madejski, G.\ 2000, ApJ, 534, 109 
\bibitem[Sikora et al.(2009)]{2009ApJ...704...38S} Sikora, M., Stawarz, {\L}., Moderski, R., Nalewajko, K., \& Madejski, G.~M.\ 2009, ApJ, 704, 38 
\bibitem[Takahashi et al.(2007)]{2007PASJ...59S..35T} Takahashi, T., et 
al.\ 2007, PASJ, 59, 35 
\bibitem[Tanaka et al.(2011)]{2011ApJ...733...19T} Tanaka, Y.~T., Stawarz, 
{\L}., Thompson, D.~J., et al.\ 2011, ApJ, 733, 19 
\bibitem[Ubertini et al.(2003)]{2003A&A...411L.131U} Ubertini, P., et al.\ 2003, A\&A, 411, L131 
\end{thebibliography}
\end{document}